\documentclass[twocolumn]{emulateapj}
\usepackage{lscape}

\slugcomment{Accepted for publication in PASP}

\shorttitle{Closure Phase of Planet Transit Events}

\shortauthors{van Belle}

\begin{document}


\title{Closure Phase Signatures of Planet Transit Events}


\author{Gerard T. van Belle\altaffilmark{1}}
\affil{European Southern Observatory, Garching, Germany 85748\\
gerard.van.belle@eso.org}


\altaffiltext{1}{For preprints, please email gerard.van.belle@eso.org.}


\begin{abstract}

Planet transit events present as attractive targets for the ultra-high-resolution capabilities afforded by optical interferometers.  Herein is presented an evaluation of the possibility of detection of such events through measurement of high-precision closure phases with the MIRC instrument on the CHARA Array.  Recovery of the transit position angle upon the sky appears readily achievable with the existing capabilities of the instrument, along with characterization of other system parameters, such as stellar radius, planet radius, and other parameters of the transit event.  This technique is the only one presently available that can provide a transiting planet's orbital plane position angle, and can directly determine the planet's radius independent of any outside observations, appearing able to improve substantially upon other determinations of that radius.  Additional directly observed parameters - also not dependent upon transit photometry or spectroscopy - include impact parameter, transit ingress time, transit velocity and stellar radius.

\end{abstract}




\section{Introduction}

Recent discoveries of stars exhibiting the telltale signs of
planet transits has begun to add a new layer of understanding to
the rapidly developing field of exoplanetology.  While the
technique of radial velocity detection has produced the greatest
yield to date of planet detections \citep{2006ApJ...646..505B}, the detected transit
events have served to define the specific nature of those planets,
including parameters such as density, atmospheric composition, and
aspects of system dynamics \citep{2006ApJ...650.1140B}.

Advances in the state of the art in astronomical optical interferometry
can be directed at these recent transit discoveries and also contribute
to filling in the pieces of the exoplanet puzzle.  Specifically, measurements
of interferometric closure phase during a planet transit event can determine the inclination and
orientation of the planetary
orbit upon the sky, in addition to refining the angular diameter measurements
of both the planet and the star.
Just as the Rossiter-McLaughlin effect \citep{1924ApJ....60...15R,1924ApJ....60...22M}
in radial velocity measurements can contribute to our knowledge of
transiting planet system parameters
\citep{2006ApJ...653L..69W},
transit event closure phases can further the physical description of
these systems through direct detection of the transit.  Interferometric phase in general
is a powerful tool that is beginning to be exploited to its fullest potential
in astronomy \citep{2007NewAR..51..604M}.

Planet transit closure phase observations described herein are the only presently available technique that provide a measurement of the transiting planet's orbital plane orientation upon the sky.  These closure phase observations also uniquely determine the other observables of the system - impact parameter, transit velocity, stellar radius, planet radius, transit ingress time - without the need for supporting observations such as transit photometry.  For example, the previous direct determination of HD189733b's diameter \citep{baines2007} measured that parameter through a combination of interferometric measurements and transit photometry; this technique is independent of such outside measurements.

We will begin with a review of instrument capabilities in \S \ref{sec_instruments}, an examination of
known extrasolar planet candidates in \S \ref{sec_potential_targets}, a `quick-and-dirty' partial analytic solution of the
problem in \S \ref{sec_partial_analytic}, a more thorough discussion of closure phase
leading to a numeric model and analysis of planet transits in \S \ref{sec_deviations}, and finally a full
Monte Carlo simulation to recover synthetic transits in \S \ref{sec_modeling}.

\section{Instrument Capabilities}\label{sec_instruments}

The Georgia State University Center for High Angular Resolution Astronomy (CHARA) Array
is a six-element optical interferometer located atop Mount Wilson in southern California.
The CHARA Array consists of six 1-m telescopes laid out on a `Y' array, two telescopes
per arm, with a baselines of $>300$ meters on the 3 longest baselines.  Initial
science operations and a facility description can be found in
\citet{2005ApJ...628..439M} and \citet{2005ApJ...628..453T}.

Commissioning on-sky tests have recently begun at the CHARA Array with the
Michigan Infrared Combiner (MIRC), a high-precision multi-telescope beam combiner
\citep{2006SPIE.6268E..55M}.  MIRC's capability to combine 4 or 6 telescopes
simultaneously and provide 3 to 10 closure phase measurements on sources represents a major
step forward in capability for the facility.  Moreover, initial MIRC tests
are indicating a remarkable ability to measure closure phases with
precision unprecedented in the field of optical interferometry,
at a level of $\sigma_\Phi \sim 0.03^o$
\citet{2006SPIE.6268E..55M}; the first science demonstrated by this capability includes
direct imaging the surface of the rapidly rotating star Altair \citep{2007Sci...317..342M}.
Other instruments, such as the VLTI AMBER instrument \citep{2006SPIE.6268E..53R},
also provide the capability to
make closure phase measurements, although initial indications are that
the closure phase precision of AMBER is not quite as capable as MIRC,
with $\sigma_\Phi \sim $few degrees \citep{2007NewAR..51..724W}.

\section{Potential Planet Transit Targets}\label{sec_potential_targets}

\begin{figure*}
\begin{center}
\includegraphics[scale=0.66,angle=0]{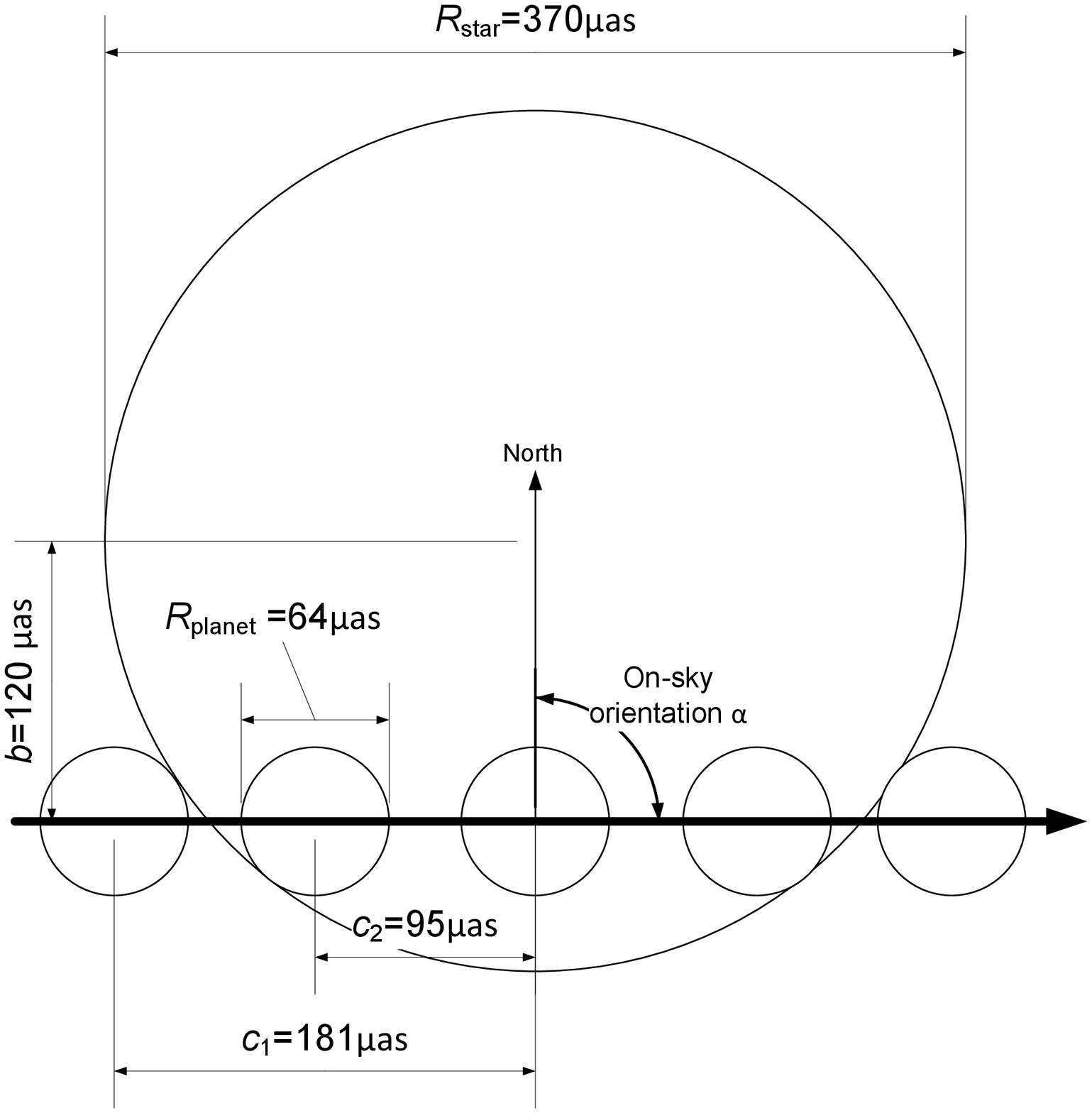}
\end{center}
\caption{\label{fig_HD189733geometry} On-sky geometry for a model representative of HD189733, based on
data found in \citet{2005A&A...444L..15B} and \citet{baines2007},
as discussed in \S \ref{sec_potential_targets}.
In particular, the planet radius of $R_{\rm planet} = 64 \mu$as and impact parameter $b = 120  \mu$as
are derived from the stellar radius, the planet-star radius ratio, and orbital inclination.
The values of $c_1 = 181$ and $c_2  = 95 \mu$as mark the distance of the beginning and end of the ingress event
from the transit meridian.
The on-sky orientation of the transit chord is, at present, unknown.}
\end{figure*}

The obvious candidate for observations of a planet transit event is
HD189733 \citep{2006ApJ...650.1160B,2005A&A...444L..15B}.
With an angular size of $376 \pm 31$ $\mu$as \citep{baines2007}, it is the
planet-transit hosting star with the largest angular size discovered to date.
The discovery paper of \citet{2005A&A...444L..15B} cites the following
system parameters: (a) a planet-star radius ratio of $0.172 \pm 0.003$,
(b) an orbital inclination of $i=85.79 \pm 0.24$, (c) an
orbital radius of $0.0313 \pm 0.0004$ AU, and (d) a transit duration of roughly 1.7 hours.
The geometry of the transit is depicted in Figure \ref{fig_HD189733geometry}.  The best
known values for the host star and planet are found in Tables 1 and 2, respectively.

\begin{deluxetable*}{lccl}
\tablecolumns{9}
\tablewidth{0pc} \tablecaption{Summary of parent star parameters for HD189733.\label{table1}}
\tablehead{
\colhead{Parameter} & \colhead{Value} & \colhead{Units} & \colhead{Reference}
}
\startdata


UD Angular Size & $366 \pm 31$ &        $\mu$as &          \citet{baines2007}  \\

Limb darkening coefficient &        1.9 &          \% & \citet{2002MNRAS.333..642T} \\

LD Angular Size & $372 \pm 31$ &        $\mu$as &         \citet{baines2007}   \\

Bolometric Flux & $28.28 \pm 0.49$  & $10^{-9}$ erg cm$^{-2}$ s$^{-1}$ &       \citet{baines2007}      \\

Parallax & $51.94 \pm 0.87$ &        mas & Perryman et al. (1997) \\

Effective Temperature & $4980 \pm  200$ &          K &            \\

Linear radius & $0.779 \pm 0.066$ &       $R_\odot$ &        \citet{baines2007}      \\

Mass & $0.82 \pm 0.05$ &       $M_\odot$ & Bouchy et al. (2005)\\

Luminosity & $0.305 \pm 0.008$ &       $L_\odot$    &            \\

log g & $4.607 \pm 0.043$ &   [cm s$^{-2}$] &            \\

\enddata
\end{deluxetable*}

\begin{deluxetable}{lccl}
\tablecolumns{9}
\tablewidth{0pc} \tablecaption{Summary of exoplanet parameters for HD189733.\label{table2}}
\tablehead{
\colhead{Parameter} & \colhead{Value} & \colhead{Units} & \colhead{Reference}
}
\startdata


Radius ratio & $0.172 \pm 0.003$ &            & Bouchy et al. 2005 \\
Angular diameter & $64.0 \pm 5.4$ &        $\mu$as &            \\
Linear Radius & $1.19 \pm 0.10$ &       Rjup &       \citet{baines2007}      \\
Density & $0.91 \pm 0.23$ &     g cm$^{-3}$ &         \citet{baines2007}    \\

\enddata
\end{deluxetable}



The next best candidate known at the time of this draft is GJ 436 \citep{2004ApJ...617..580B,2007A&A...472L..13G}, with roughly
the same anticipated angular size as HD189733.
HD149026 \citep{2005ApJ...633..465S}, HD17156 \citep{2007ApJ...669.1336F,2007A&A...476L..13B}, and HD209458 \citep{1999IAUC.7307....1H,2000ApJ...529L..45C} are also worth considering, although their
stellar angular diameters of 170-250 $\mu$as are significantly less favorable for detection
when considering the currently available capabilities of CHARA-MIRC.

\section{Partial Analytic Solution for Transit Interferometric Visibility}\label{sec_partial_analytic}

The complex interferometric visibility of a binary star can be written as:
\begin{equation}\label{eqn_Vbinary}
V_{\rm binary} = e^{-2 \pi i (u \alpha_1 + v \beta_1)}
{V_A + r V_B e^{-2 \pi i (u \Delta \alpha + v \Delta \beta)}
\over
1 + r}
\end{equation}
where $r$ is the brightness ratio \citep{1971MNRAS.151..161H}.
$V_A$ and $V_B$ are the visibility functions associated with a uniform disk, $V = 2 J_1 (x) / (x)$,
where $x=\theta_{UD} \pi B / \lambda$, $\theta_{UD}$ is the uniform disk angular diameter, $B$ is the
projected baseline, and $\lambda$ is the wavelength of operation.  Dropping the absolute phase term
and rewriting this in terms of relative separation vector $\Delta {\bf s}$ and baseline vector ${\bf B}$:
\begin{equation}\label{eqn_Vsingle}
V_{binary} =
{V_A + r V_B e^{{-2 \pi i{{\bf B} \cdot \Delta {\bf s} / \lambda}  }}
\over
1 + r}
\end{equation}

Determining visibilities for the specific case of a planet transit can
be adopted from this formalization
with the following caveats:
\begin{itemize}
\item The `brightness ratio' $r$ is the negative value of the squared planet-to-star diameter ratio.
\item Equation \ref{eqn_Vsingle} assumes that both the planet and stellar disks are indeed uniform
disks, since $r$ is constant.  Specifically, limb darkening of the star is ignored.
\item Equation \ref{eqn_Vsingle} is valid only for the portion of the transit event
when the planet is fully in front of the star, when $r$ is again constant.  The ingress and egress portions
of the transit event are not properly represented by this equation.
\end{itemize}

This solution is useful in characterizing the order-of-magnitude
effects for observation planning; however, given the above caveats,
it is insufficient for proper evaluation of actual data.

\section{Transit Closure Phase and Visibility Deviations}\label{sec_deviations}

In its simplest realization, an optical interferometer
measures the Fourier components of an image upon the sky.  The
location of the image's components in the Fourier transformed $\{u,v\}$ plane are dictated
by the baseline between the telescopes in the interferometer projected towards the source of interest, and the
wavelength of operation.  As seen in \S \ref{sec_partial_analytic},
at least a partial analytic solution can be predicted from
the parameters of the experiment.  In practice, only the visibility amplitude
(typically just referred to as the {\it visibility}) can be measured,
while atmospheric turbulence corrupts the direct measurement
of visibility phase.

However, interferometers using three or more telescopes
can produce a measurement of the {\it closure phase}, a phase quantity
that remains uncorrupted by telescope-specific phase errors
\citep{1958MNRAS.118..276J,2000plbs.conf..203M}.  The closure phase $\Phi$ is
the sum of visibility phases around a closed loop of baselines.  For
three telescopes $i,j,k$, this is easily deduced from the observed
visibilities:
\begin{equation}
\Phi_{ijk} = arg(V_{ij}) + arg(V_{jk}) + arg(V_{ki})
\end{equation}

Each pair of telescopes
produces a source visibility and phase, where the phase is associated with the source's intrinsic
phase $\phi$, phase errors $\theta_1-\theta_2$ associated with the telescope pair, and noise.  Typically
due to atmospheric corruption, for a pair of telescopes, phase information is useless.  However,
for a three telescope array $\{l,m,n\}$, combination of the three measured phase pairs
$(\psi_{lm}=\phi_{lm}+(\theta_l-\theta_m),\ldots)$ results in cancelation of the phase error terms $\theta$
leaving only the sum of the three source phases intrinsic to the object - the closure phase, $\Phi_{lmn} = \phi_{lm}+\phi_{mn}+\phi_{nl}$.

Use of the closure phase effectively cancels many of the corrupting effects of
the atmosphere and the instrument, and is a highly sensitive probe for interferometric image construction
on the smallest spatial scales.  Significantly more complete discussions of the topic of
closure phase may be found in \citet{1984ARA&A..22...97P} and \citet{2007NewAR..51..604M}.
Closure phases have been used to explore disk asymmetries
in YSOs \citep{2006ApJ...647..444M} and are very sensitive to asymmetries in
images, which will prove quite useful in the application discussed here.

For a star with a planet blocking out part of its disk during a transit event, the degree
of asymmetry is extreme - significantly much more so than a star with a starspot on its
surface: the spot temperature is merely some slight fraction of the rest of the photosphere and is
still emitting radiation at a fairly significant level.  At the near-infared wavelengths being
considered here, a transiting planet
emits extremely little radiation with regards to the area of the stellar photosphere it is blocking
off from our line of sight.

For the full envelope of expected visibility amplitudes and closure phases
for a {\it gedanken} experiment covering the interferometer response
during a planet transit event, the analytic solution of \S \ref{sec_partial_analytic}
is insufficient, in that it breaks down during the transit ingress and egress.
As a means to explore that full envelope, a numeric analysis can be performed
to compute directly the expected visibility components.

For the specific case of HD189733, such an experiment may easily be
executed, leading to an expectation of the changes in observed visibility amplitudes and
closure phases during the planet transit event.
For the {\it gedanken} experiment, three baselines were postulated, with
coordinates (in meters) of $\{125.35, 305.94\}, \{-300.42, -89.62\}, \{175.07, -216.32\}$ (with zero
vertical separation), which
correspond roughly to the CHARA E1S1, W1E1, and S1W1 baselines, respectively.

\begin{figure*}
\plotone{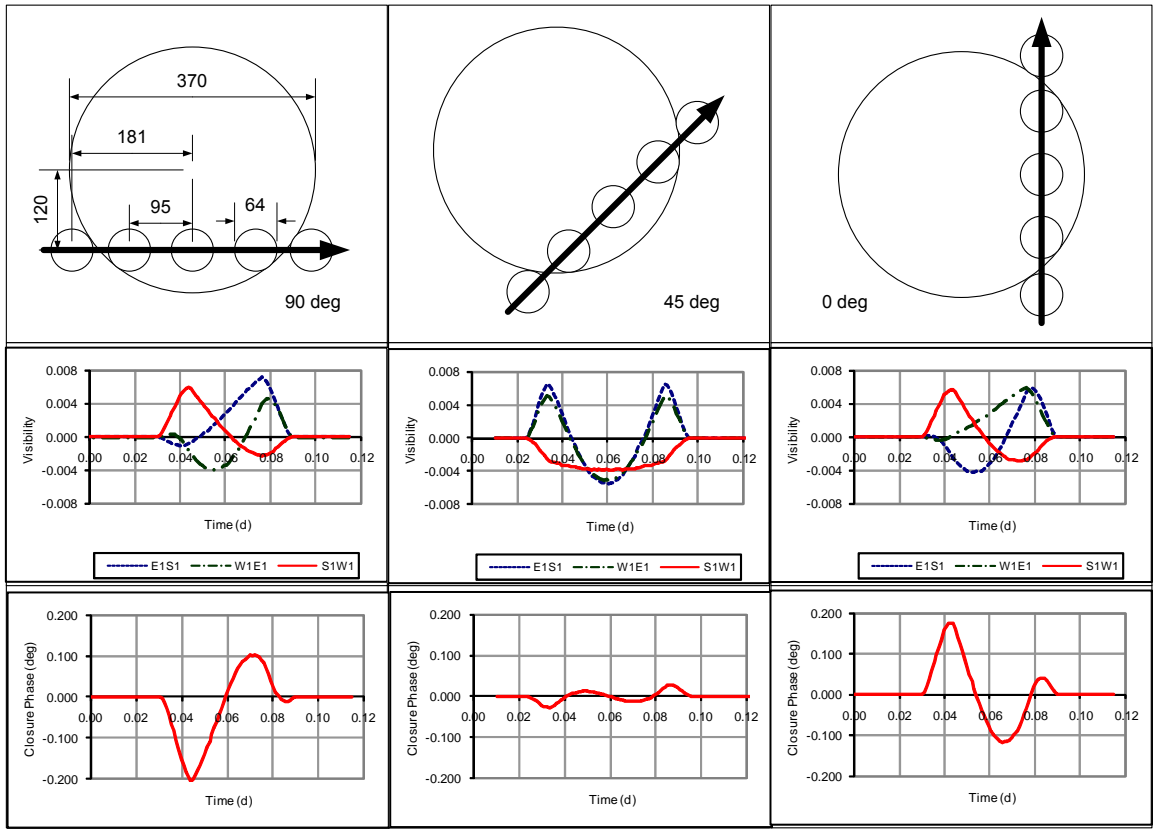}
\caption{\label{fig_HD189733CP} Excursions in visibility amplitude
and closure phase data for HD 189733 as observed by CHARA for 3 different
orientations,
as discussed in \S \ref{sec_modeling}.
The top row is the image on the sky, the middle row is the visibility amplitude excursions
for each of the 3 CHARA baselines discussed herein, and the bottom row is the
closure phase difference during the transit event.}
\end{figure*}

From the system parameters for HD189733 as cited by \citet{2005A&A...444L..15B},
a model of the transit event was constructed with a stellar radius of 185$\mu$as,
a planet radius of 32$\mu$as, and a transit chord that was offset from the center of
the stellar disk by 121$\mu$as.  The rate of the transit was not considered in this section,
in this simple inspection of the effects upon the interferometer visibility signals,
although in \S \ref{sec_modeling} variables for that aspect of the event will be introduced.
The top panels of Figure \ref{fig_HD189733CP} shows the this transit event for three different orientations upon the sky,
$\alpha = 0^o,45^o,90^o$.

For each of the three orientations, the visibility for each of the three baselines, along
with the closure phase, was computed.  This computation was performed in the following manner:
An appropriately limb darkened
model star was created numerically on a $1024 \times 1024$ grid with a diameter corresponding
to the HD189733 parent star.
As introduced by \citet{1921MNRAS..81..361M} and discussed in the
context of stellar interferometry by
\citet{1974MNRAS.167..475B}, the conventional linear representation
of limb darkening across the disk of a star can be written as:
\begin{equation}
I_\lambda(\mu) = I_\lambda(1)[1-u_\lambda(1-\mu)]
\end{equation}
where $\mu=\cos \gamma$, $\gamma$ is the angle between the line of
sight and the stellar surface normal, and $I_\lambda(1)$ is the
specific intensity at the center of the disk. For HD189733 at $5000\pm100$K
\citep{baines2007}, the limb darkening parameter $u$ at 1.6 $\mu$m is roughly
equal to 0.35 \citep{1995A&AS..114..247C}; as we shall see in \S \ref{sec_modeling},
this technique is relative insensitive
to limb darkening of the planet host star.  This latter fact is unsurprising
given the small angular size of the star relative to our notional
array; for an array with larger baselines ($B\sim 1$km), there would a
greater sensitivity to this parameter.

A fully darkened spot corresponding to the radius of the transiting planet was then created on the
image for a given location along that transit, and the Fourier components were computed.
Rather than bear the full computational load of Fourier transforming the entire image
upon the sky, the approach of \citet{2006ApJ...645..664A} is followed, and
only the specific components corresponding to the 3 baselines in question were computed,
resulting in a much lighter computational load without a sacrifice in precision.

This process was then repeated for various points along the planet transit.  The computed values
for visibility and closure phase were
compared to the nominal values for the uneclipsed parent star, and
those deviations are plotted in the middle and lower panel of Figure \ref{fig_HD189733CP}.
This approach to computing the visibility amplitudes and closure phases
can be seen as being superior to the analytic solution in \S \ref{sec_partial_analytic},
since it takes into account stellar limb darkening, and is valid through the transit
event, including ingress and egress.

As seen in Figure \ref{fig_HD189733CP}, during the transit event,
the visibility deviates from from the nominal unocculted star case, but
by only a marginal amount - on the order of $\pm0.01\%$.  Such a measurement is
beyond the capabilities of any existing interferometer by two orders of magnitude.  However, the
closure phase excursion is $\pm0.2^o$.  As detailed in \citet{2006SPIE.6268E..55M}, closure phase
measurements at this level of precision appear possible: initial tests of the CHARA-MIRC
system showed closure phase formal error at the $\sigma_\Phi \sim 0.03^o$ level over the course of 3 hours.
Shorter integration times indicated a correspondingly higher level of scatter, but the magnitude
of this error gives a starting point from which to evaluate the possibility for observation
of a planet transit event using closure phases.

\section{Recovery of Transit Parameters from Closure Phase Measurements}\label{sec_modeling}

Having demonstrated in \S \ref{sec_deviations} in at least a qualitative way that
closure phase excursions result from a planet transit event, it is useful to
further demonstrate that an event can be reconstructed from an ensemble of
closure phase measurements taken during a transit.
Free parameters of the fit are the primary descriptors of the image
upon the sky:
\begin{itemize}
\item The stellar radius $r_{star}$.
\item The planet to star radius ratio $R$.
\item The orbit orientation upon the sky $\alpha$, defined as the angular orientation of the chord of the transit event
across the disk of the star in right ascension and declination, as measured from north to east on the sky.
\item The `impact parameter' $b$, defined as the distance between the chord of the transit event across the
disk of the star, and the center of the stellar disk.
\item The `zero time' $JD_0$ of the transit event, defined here as the time of closest approach of
the planet disk to the center of the stellar disk.
\item The velocity $v$ of the planet disk across the disk of the star.
\end{itemize}

Fixed input parameters of the fit are:
\begin{itemize}
\item The $i = 1 \dots N$ closure phases $\Phi_i$ and their associated errors $\sigma_{\Phi,i}$.
For this exercise, it will be assumed that these are the single closure phases associated with the
non-diurnally evolved CHARA baselines associated with the S1, S2, E1, and W1 stations.  In practice, CHARA-MIRC is operated with
either four or six simultaneous baselines, which in the latter case can potentially provide even more closure phases than
being modelled here; this, in conjunction with baseline diurnal evolution, would serve to further
constrain the transit event parameters.
\item The closure phase errors $\sigma_{\Phi,i}$.  These errors were assumed to have a normal distribution,
and for the various synthetic data sets created below, assumed to be of a magnitude ranging from $\overline{\sigma_{\Phi,i}}=0.005^o, 0.010^o, 0.020^o, 0.050^o, 0.100^o$.  The
FWHM of the $\sigma_{\Phi,i}$ distribution was set at one-quarter of the average error value.  Individual
closure phases $\Phi_i$ were randomized by an error of the scope of $\sigma_{\Phi,i}$ (but independent thereof).
\item The operational wavelength $\lambda_i$ of each observation of the closure phases.  A single
wavelength of 1.6 $\mu$m will be assumed here for the synthetic data sets.  In practice, the MIRC instrument spectrally
disperses the starlight and multiple closure phases per observation are available as a result.
\item The time $t_i$ of observation of each closure phase. For our synthetic data sets, we will postulate that sets of
closure phases are taken sequentially
at intervals of 0.001 of a Julian Day (roughly 85 seconds set-to-set), for a total duration of 0.124 days (about
3 hours), which spans the duration of the transit event, plus about 40 minutes before and after the transit
event.  This resulted in $N=124$ data points per observation set.
\item The $\{u,v\}$ coordinates of observing baselines for each closure phase,
For an actual observation, the baselines would need to be properly projected onto
the sky incorporating diurnal motion.
\item The limb darkening of the star.
\end{itemize}
The stellar limb darkening is also potentially
fittable free parameter; however, for the marginal resolution case of
CHARA observing HD189733, a model value is sufficient for the
fitting.

For simplicity of this investigation, diurnal motion will be ignored and the $\{u,v\}$ coordinates
of the observing baselines will be constant.  In practice, this motion will need to be accounted
for but will actually provide addition constraints upon the image reconstruction, much in the
same way that baseline evolution can serve to assist in constraining the parametrization of
binary star orbits \citep{1999ApJ...515..356B}.

To create synthetic `observation sets' for testing our fitting and parameter recovery routines,
the free parameters $\{r_{star},R,b,{JD}_0,v\}$ discussed at the beginning of this section were set to the values seen in our
HD189733-like system in Figure \ref{fig_HD189733geometry} (with $JD_0=-6000$ sec, $R=0.17$, $b=121.0 \mu$as, $r_{\rm star}=185 \mu$as, and $v=0.07000 \mu$as/sec), and data sets were generated within the context of the fixed input parameters discussed above.  For those data sets, the position
angle $\alpha$ of the planet transit across the stellar disk was also set to
the values $28^o$, $115^o$, and $170^o$, to test
the sensitivity of the parameter recovery on that particular parameter as well.

Thus, to test 'goodness of fit' for a given set of six randomized free parameters
$\{r_{star},R,\alpha,b,JD_0,v\}$, a model transit event sequence was generated, projected upon the sky,
and resultant image sequence Fourier transformed for
comparison to each of the observed $\Phi_i$ data points,
and a $\chi^2/$DOF calculated.  A
multi-dimensional optimization code was then utilized to loop about this goodness of fit routine and derive
the best $\{r_{star},R,\alpha,b,JD_0,v\}$ solution from any given
starting point, a process that took typically 500 iterations
\citep{pre92}.
An exhaustive search of the transit event parameter space was used
to explore the $\chi^2/$DOF space, using a two-fold approach.
First, a grid of reasonable $\{r_{star},R,\alpha,b,JD_0,v\}$ starting values was explored to see if the original $\{r_{star},R,\alpha,b,JD_0,v\}$
parameters could be recovered, with $\alpha$ varying over its full range of $\{0^o,360^o\}$, and the other parameters being explored over a range of $\pm 50\%$ of their `true' values.  These latter ranges were expected to encompass the reasonable starting points for an actual investigation, based upon constraints that may be available from discovery photometry or spectroscopy.  Each range was gridded with a density of 5 to 10 points per variable.  Second, a large number of iterations $(N\sim1000)$ were also run for each
synthetic data set starting from fully randomized $\{r_{star},R,\alpha,b,JD_0,v\}$ starting values, also with the purpose of
recovering the original $\{r_{star},R,\alpha,b,JD_0,v\}$
parameters that described the generating values of the synthetic data set.  The resulting $\chi^2$ manifold appeared
to be smoothly varying, as the recovery of the original parameters appeared to occur without local minima obstructing
the recovery of the global minima.
Once the best solution was established for a given data set, $1-\sigma$ errors
were established about the $\chi^2$ minimum through exploration of appropriate $\Delta \chi^2$ intervals.
The results for each of the three transit position angles,
with the five different levels of closure phase error, are seen in Table \ref{table3}.

\section{Discussion and Conclusion}

Examination of Table \ref{table3} illustrates that, even with a crude level of closure phase error ($\overline{\sigma_\Phi} \sim 0.1^o$), the position angle of the transit event is readily recovered to within a few degrees, and markedly better with modest improvements in closure phase error.  Since closure phase is, in essence, an observable that quantifies the degree of asymmetry in an image upon the sky, we expect that this technique should work well even for grazing transit events.  These events would be limited presumably by the shorter duration of the occultation event (and thus fewer closure phase data points to fit), and also by a lesser stellar surface area occulted by the planet's disk, resulting in a smaller closure phase signal, but the basic promise of the approach holds true.  The sensitivity of this approach to limb darkening - mentioned, but largely dismissed in \S \ref{sec_modeling} - is only slight, due to the largely axisymmetric nature of limb darkening.

Of particular interest is the fact that, in the best conceivable case for each apparent transit position angle, the planet radius appears to be recoverable to the level of roughly one part in 30-40, which appears to best the previous interferometric measure of the planetary diameter \citep{baines2007} by a factor of 2.5.  Additionally, the technique establishes that diameter in a way that is independent of the transit photometry - the \citet{baines2007} investigation relied upon transit photometry for a value of the planet-star radius, $R$.  There is further the possibility that conducting such an observation in a wavelength-dependent sense could probe molecular opacity effects of the the planetary atmosphere though sensing apparent radius dependencies.  For example, the recent detection of methane as an constituent of HD189733b's atmosphere \citep{2008arXiv0802.1030S} provides a tantalizing goal for this technique; in principle, a sufficiently precise measurement of this nature, using narrow channels inside of the $H$-band (say, comparing 0.05 $\mu$m-wide channels centered at 1.5, 1.66, and 1.73 $\mu$, respectively) could confirm this detection by detecting the wavelength dependence of the planetary radius.
However, this appears to require levels of closure phase precision beyond even the best cases considered here - an examination of the absorption depth data of \citet{2008arXiv0802.1030S} indicates such a detection to require radius measurement precision at the $0.2 - 1.1 \%$ level.

These simulations use {\it only} the time-tagged closure phase data from a single transit event; supporting photometric and radial velocity signatures of the transit (and interferometer visibility measures), or multiple transits, have the potential to significantly improve the quality of the $\{r_{star},R,\alpha,b,JD_0\}$ fit parameters.  Investigators wishing to mesh such data sets will of course have to pay particular attention to uniform time-tagging.  It does seem possible, however, that data sets of such richness will be able to probe other system parameters: possible moons of the transiting planet, and the presence of other stellar planetary companions due to variations in the transit timing and impact parameter.

This approach is the {\bf only} currently available technique that provide any value for the transit event orientation angle, $\alpha$; it is also an independent check on parameters such as stellar radius or planet-star radius ratio that is derived from other techniques, such as spectroscopy or photometric timing.  Use of results from such interferometric observations could be highly useful for planning observations of TPF-I, Darwin, and other instruments that have a position-angle dependent response.  For example, the extreme adaptive optics systems that are envisioned carrying out planet searches and/or characterizations through `dark hole' techniques \citep{2007ApJ...658.1386S,2008arXiv0803.3629O} could have their search times reduced through {\it a priori} knowledge of the planet's orbital plane position angle.

\acknowledgements

Special thanks to Theo ten Brummelaar, John Monnier, Mark Swain, and Ming Zhao for particularly helpful feedback during the development of this manuscript, which also benefited greatly from the input of an anonymous referee. The CHARA Array is funded by the National Science Foundation through grant AST 94-14449, the W. M. Keck Foundation, the David and Lucile Packard Foundation, and by Georgia State University. This research has made use of the SIMBAD literature database, operated at CDS, Strasbourg, France, the FUTDI database at AMNH, and of NASA's Astrophysics Data System.

\clearpage
\begin{landscape}

\begin{deluxetable}{ccccccccccc}
\tablecolumns{11}
\tabletypesize{\scriptsize}
\tablewidth{0pc} \tablecaption{Results from fitting transit event parameters to three synthetic transit closure phase data sets as discussed in \S \ref{sec_modeling}.  Original values were $JD_0=-6000$ sec, $R=0.17$, $b=121.0 \mu$as, $r_{\rm star}=185 \mu$as, and $v=0.07000 \mu$as/sec.\label{table3}}
\tablehead{
 \colhead{Input} &   & \colhead{Time}
& \colhead{Ratio}	& \colhead{Impact}	& \colhead{Recovered}	& \colhead{Star}	& \colhead{Planet}		& \colhead{Planet}	& \colhead{Planet Radius}	\\
\colhead{Position Angle} & \colhead{$\overline{\sigma_{\Phi}}$} & \colhead{Zero}
& \colhead{Planet-Star}	& \colhead{Parameter}	& \colhead{Position Angle}	& \colhead{Radius}	& \colhead{Speed}		 & \colhead{Radius}	& \colhead{Fractional}	\\
\colhead{($\alpha$) [deg]} & \colhead{[deg]} & \colhead{($JD_0$) [sec]} & \colhead{Radius $(R)$} & \colhead{$(b)$ [$\mu$as]} & \colhead{$(\alpha)$ [deg]} & \colhead{$(r_{\rm star})$ [$\mu$as]} & \colhead{$(v)$ [$\mu$as/sec]}  & \colhead{$(r_{\rm planet})$ [$\mu$as]} & \colhead{Error}
}
\startdata

28 & 0.005 & $-5981 \pm 20$ & $0.1709 \pm 0.0061$ & $119.9 \pm 2.6$ & $27.87 \pm 0.26$ & $185.2 \pm 1.7$ & $0.07000 \pm 0.00047$ & $31.7 \pm 1.17$ & $27.1$ \\
28 & 0.010 & $-6075 \pm 38$ & $0.1701 \pm 0.0074$ & $123.2 \pm 2.2$ & $28.25 \pm 0.42$ & $185.5 \pm 2.7$ & $0.06900 \pm 0.00092$ & $31.6 \pm 1.45$ & $21.8$ \\
28 & 0.020 & $-5969 \pm 53$ & $0.1687 \pm 0.0115$ & $114.1 \pm 5.9$ & $27.02 \pm 0.82$ & $183.2 \pm 4.3$ & $0.06900 \pm 0.00180$ & $30.9 \pm 2.23$ & $13.9$ \\
28 & 0.050 & $-6268 \pm 97$ & $0.1717 \pm 0.0126$ & $137.1 \pm 6.3$ & $27.67 \pm 1.40$ & $189.1 \pm 7.3$ & $0.07100 \pm 0.00352$ & $32.5 \pm 2.69$ & $12.1$ \\
28 & 0.100 & $-5917 \pm 183$ & $0.1748 \pm 0.0352$ & $124.5 \pm 12.2$ & $25.98 \pm 2.98$ & $183.3 \pm 18.2$ & $0.07000 \pm 0.00640$ & $32.0 \pm 7.19$ & $4.5$ \\
115 & 0.005 & $-5955 \pm 28$ & $0.1719 \pm 0.0047$ & $123.1 \pm 1.2$ & $115.52 \pm 0.37$ & $184.3 \pm 1.9$ & $0.07000 \pm 0.00097$ & $31.7 \pm 0.93$ & $34.2$ \\
115 & 0.010 & $-6020 \pm 21$ & $0.1690 \pm 0.0038$ & $120.3 \pm 1.8$ & $114.95 \pm 0.42$ & $185.3 \pm 1.5$ & $0.07000 \pm 0.00095$ & $31.3 \pm 0.75$ & $41.8$ \\
115 & 0.020 & $-6011 \pm 41$ & $0.1700 \pm 0.0075$ & $120.1 \pm 3.4$ & $114.84 \pm 0.85$ & $184.4 \pm 2.4$ & $0.06900 \pm 0.00143$ & $31.3 \pm 1.44$ & $21.7$ \\
115 & 0.050 & $-5874 \pm 77$ & $0.1708 \pm 0.0121$ & $123.9 \pm 5.6$ & $115.37 \pm 1.70$ & $180.3 \pm 5.7$ & $0.06800 \pm 0.00321$ & $30.8 \pm 2.39$ & $12.9$ \\
115 & 0.100 & $-5957 \pm 155$ & $0.1742 \pm 0.0178$ & $136.2 \pm 12.8$ & $115.34 \pm 2.64$ & $184.7 \pm 8.1$ & $0.06900 \pm 0.00507$ & $32.2 \pm 3.58$ & $9.0$ \\
170 & 0.005 & $-6005 \pm 12$ & $0.1694 \pm 0.0051$ & $121.4 \pm 1.6$ & $170.13 \pm 0.31$ & $185.6 \pm 1.8$ & $0.07100 \pm 0.00121$ & $31.4 \pm 0.99$ & $31.6$ \\
170 & 0.010 & $-6001 \pm 17$ & $0.1703 \pm 0.0053$ & $121.8 \pm 2.1$ & $170.05 \pm 0.44$ & $183.8 \pm 1.5$ & $0.06900 \pm 0.00065$ & $31.3 \pm 1.01$ & $31.1$ \\
170 & 0.020 & $-5920 \pm 36$ & $0.1698 \pm 0.0076$ & $120.2 \pm 4.5$ & $168.51 \pm 0.67$ & $181.0 \pm 2.9$ & $0.06800 \pm 0.00182$ & $30.7 \pm 1.46$ & $21.0$ \\
170 & 0.050 & $-5943 \pm 65$ & $0.1715 \pm 0.0193$ & $122.0 \pm 6.6$ & $168.04 \pm 1.31$ & $186.7 \pm 5.3$ & $0.06900 \pm 0.00281$ & $32.0 \pm 3.72$ & $8.6$ \\
170 & 0.100 & $-5893 \pm 236$ & $0.1696 \pm 0.0275$ & $123.7 \pm 13.6$ & $168.12 \pm 4.87$ & $178.1 \pm 13.3$ & $0.06400 \pm 0.00756$ & $30.2 \pm 5.39$ & $5.6$ \\

\enddata
\end{deluxetable}



\clearpage
\end{landscape}


\begin{thebibliography}{}

\bibitem[Aufdenberg et al.(2006)]{2006ApJ...645..664A} Aufdenberg, J.~P., et al.\ 2006, \apj, 645, 664
\bibitem[Baines et al.(2007)]{baines2007} Baines, E.~K., van Belle, G.~T., ten Brummelaar, T.~A., McAlister, H.~A., Swain, M., Turner, N.~H., Sturmann, L., \& Sturmann, J.\ 2007, \apjl, 661, L195
\bibitem[Bakos et al.(2006)]{2006ApJ...650.1160B} Bakos, G.~{\'A}., et al.\ 2006, \apj, 650, 1160
\bibitem[Barbieri et al.(2007)]{2007A&A...476L..13B} Barbieri, M., et al.\ 2007, \aap, 476, L13
\bibitem[Boden et al.(1999)]{1999ApJ...515..356B} Boden, A.~F., et al.\ 1999, \apj, 515, 356
\bibitem[Bouchy et al.(2005)]{2005A&A...444L..15B} Bouchy, F., et al. 2005, \aap, 444, L15
\bibitem[Butler et al.(2004)]{2004ApJ...617..580B} Butler, R.~P., Vogt, S.~S., Marcy, G.~W., Fischer, D.~A., Wright, J.~T., Henry, G.~W., Laughlin, G., \& Lissauer, J.~J.\ 2004, \apj, 617, 580
\bibitem[Butler et al.(2006)]{2006ApJ...646..505B} Butler, R.~P., et al.\ 2006, \apj, 646, 505
\bibitem[Burrows et al.(2006)]{2006ApJ...650.1140B} Burrows, A., Sudarsky, D., \& Hubeny, I.\ 2006, \apj, 650, 1140
\bibitem[Charbonneau et al.(2000)]{2000ApJ...529L..45C} Charbonneau, D., Brown, T.~M., Latham, D.~W., \& Mayor, M.\ 2000, \apjl, 529, L45
\bibitem[Claret et al.(1995)]{1995A&AS..114..247C} Claret, A., Diaz-Cordoves, J., \& Gimenez, A.\ 1995, \aaps, 114, 247
\bibitem[Cody \& Sasselov(2002)]{2002ApJ...569..451C} Cody, A.~M., \& Sasselov, D.~D.\ 2002, \apj, 569, 451
\bibitem[Fischer et al.(2007)]{2007ApJ...669.1336F} Fischer, D.~A., et al.\ 2007, \apj, 669, 1336
\bibitem[Gillon et al.(2007)]{2007A&A...472L..13G} Gillon, M., et al.\ 2007, \aap, 472, L13
\bibitem[Hanbury Brown et al.(1974)]{1974MNRAS.167..475B} Hanbury Brown, R., Davis, J., Lake, R.~J.~W., \& Thompson, R.~J.\ 1974, \mnras, 167, 475
\bibitem[Henry et al.(1999)]{1999IAUC.7307....1H} Henry, G.~W., Marcy, G., Butler, R.~P., \& Vogt, S.~S.\ 1999, \iaucirc, 7307, 1
\bibitem[Herbison-Evans et al.(1971)]{1971MNRAS.151..161H} Herbison-Evans, D., Hanbury Brown, R., Davis, J., \& Allen, L.~R.\ 1971, \mnras, 151, 161
\bibitem[Jennison(1958)]{1958MNRAS.118..276J} Jennison, R.~C.\ 1958, \mnras, 118, 276
\bibitem[McLaughlin(1924)]{1924ApJ....60...22M} McLaughlin, D.~B.\ 1924, \apj, 60, 22
\bibitem[Milne(1921)]{1921MNRAS..81..361M} Milne, E.~A.\ 1921, \mnras, 81, 361
\bibitem[McAlister et al.(2005)]{2005ApJ...628..439M} McAlister, H.~A., et al.\ 2005, \apj, 628, 439
\bibitem[Monnier(2000)]{2000plbs.conf..203M} Monnier, J.~D.\ 2000, Principles of Long Baseline Stellar Interferometry, 203
\bibitem[Monnier et al.(2006a)]{2006SPIE.6268E..55M} Monnier, J.~D., et al.\ 2006a, \procspie, 6268
\bibitem[Monnier et al.(2006b)]{2006ApJ...647..444M} Monnier, J.~D., et al.\ 2006b, \apj, 647, 444
\bibitem[Monnier et al.(2007)]{2007Sci...317..342M} Monnier, J.~D., et al.\ 2007, Science, 317, 342
\bibitem[Monnier(2007)]{2007NewAR..51..604M} Monnier, J.~D.\ 2007, New Astronomy Review, 51, 604
\bibitem[Oppenheimer et al.(2008)]{2008arXiv0803.3629O} Oppenheimer, B.~R., et al.\ 2008, ArXiv e-prints, 803, arXiv:0803.3629
\bibitem[Pearson \& Readhead(1984)]{1984ARA&A..22...97P} Pearson, T.~J., \& Readhead, A.~C.~S.\ 1984, \araa, 22, 97
\bibitem[Press et al.(1992)]{pre92}Press, W.H., Teukolsky, S.A., Vetterling, W.T., Flannery, B.P., 1992, Numerical Receipes in C, Port Chester, Cambridge University Press
\bibitem[Rantakyr{\"o} et al.(2006)]{2006SPIE.6268E..53R} Rantakyr{\"o}, F.~T., et al.\ 2006, \procspie, 6268
\bibitem[Rossiter(1924)]{1924ApJ....60...15R} Rossiter, R.~A.\ 1924, \apj, 60, 15
\bibitem[Sato et al.(2005)]{2005ApJ...633..465S} Sato, B., et al.\ 2005, \apj, 633, 465
\bibitem[Serabyn et al.(2007)]{2007ApJ...658.1386S} Serabyn, E., Wallace, K., Troy, M., Mennesson, B., Haguenauer, P., Gappinger, R., \& Burruss, R.\ 2007, \apj, 658, 1386
\bibitem[Swain et al.(2008)]{2008arXiv0802.1030S} Swain, M.~R., Vasisht, G., \& Tinetti, G.\ 2008, ArXiv e-prints, 802, arXiv:0802.1030
\bibitem[Tango \& Davis(2002)]{2002MNRAS.333..642T} Tango, W.~J., \& Davis, J.\ 2002, \mnras, 333, 642
\bibitem[ten Brummelaar et al.(2005)]{2005ApJ...628..453T} ten Brummelaar, T.~A., et al.\ 2005, \apj, 628, 453
\bibitem[Weigelt et al.(2007)]{2007NewAR..51..724W} Weigelt, G., Driebe, T., Hofmann, K.-H., Kraus, S., Petrov, R., \& Schertl, D.\ 2007, New Astronomy Review, 51, 724
\bibitem[Winn et al.(2006)]{2006ApJ...653L..69W} Winn, J.~N., et al.\ 2006, \apjl, 653, L69


\end{thebibliography}
\end{document}